# Emergent Inductance from Chiral Orbital Currents in a Bulk Ferrimagnet

Gang Cao[1,2*], Hengdi Zhao[1], Yu Zhang[1], Alex Fix[1], Tristan R. Cao[1], Dhruva Ananth[1], Yifei Ni[1], Gabriel Schebel[1], Rahul Nandkishore[1,3], Itamar Kimchi[4], Hua Chen[5], Feng Ye[6] and Lance E. DeLong[7]

[1]*Department of Physics, University of Colorado at Boulder, Boulder, CO 80309*

[2]*Center for Experiments on Quantum Materials, University of Colorado at Boulder, Boulder, CO 80309*

[3]*Center for Theory of Quantum Matter, University of Colorado at Boulder, Boulder, CO 80309*

[4]*School of Physics, Georgia Institute of Technology, Atlanta, GA 30332, USA*

[5]*Department of Physics and Astronomy, Colorado State University, Fort Collins, CO 80521*

[6]*Neutron Scattering Division, Oak Ridge National Laboratory, Oak Ridge, Tennessee 37831*

[7]*Department of Physics and Astronomy, University of Kentucky, Lexington, KY 40506*

We report the discovery of a new form of inductance in the bulk ferrimagnet $Mn_3Si_2Te_6$, which features strong spin-orbit coupling, large magnetic anisotropy, and pronounced magnetoelastic interactions. Below its Curie temperature ($T_C \approx 78$ K), $Mn_3Si_2Te_6$ hosts chiral orbital currents (COC) that circulate within the crystal lattice and give rise to collective electronic behavior [1]. By applying a magnetic field along the hard *c* axis and driving the system with low-frequency currents, we uncover a giant inductive response up to millihenry scale, originating from first-order reconfigurations of COC domains. These domains act as coherent mesoscopic inductive elements that resist reversal upon current reduction, producing a large electromotive force and sharply increasing voltage. This emergent inductance defies classical models, occurs without superconductivity or engineered nanostructures, and opens a new frontier in orbital-based quantum functionality and device concepts.

*Corresponding author: gang.cao@colorado.edu

Inductance is a cornerstone concept in electromagnetism, governing the response of materials and devices to time-varying currents. Traditionally, two primary forms of inductance are recognized. *Geometrical inductance* arises from the magnetic flux generated by current loops, as described by Faraday's and Lenz's laws, and is dictated by the physical geometry of conductive pathways. *Kinetic inductance*, in contrast, results from the inertia of charge carriers and becomes significant in superconductors, where electron scattering is minimal. More recently, a third form of inductance has been identified in nanoscale systems with helical-spin textures, where the Berry phase acquired by spin-aligned electrons induces an emergent electromotive force [2-5]. Despite these advances, all known forms of inductance rely either on macroscopic magnetic structures or superconducting coherence, limiting their applicability and tunability in bulk materials.

In this work, we report the discovery of a fundamentally new inductive phenomenon in bulk single crystals of the ferrimagnet $Mn_3Si_2Te_6$ [1, 6-17] (See Supplemental Material [17] attached to the end of this paper for experimental details including thermal diagnostics *conclusively eliminating Joule heating as an origin of reported novel phenomena* [1, 6-9]). This effect arises from *chiral orbital currents* (COC) circulating at the atomic scale, stabilized by strong spin-orbit interactions (SOI) and magnetoelastic coupling [1]. Unlike conventional inductance mechanisms, this inductive response emerges without the need for superconductivity or engineered nanoscale structures. Under a constant magnetic field aligned along the hard magnetic *c*-axis and a slowly varying current, the COC undergo first-order reconfigurations into coherent domains that resist reversal when the current, *I*, decreases. This results in a dramatic increase in voltage, V, by up to three orders of magnitude as *I* is reduced, manifesting an exceptionally large inductive effect that defies classical models. The existence of such a phenomenon in a bulk material introduces a new paradigm for understanding electromagnetic responses driven by intertwined orbital, spin, and lattice dynamics. The slow dynamics of the COC reconfigurations, spanning seconds to minutes [1], further distinguish this state from conventional fast electronic processes. Our findings open new avenues for exploring emergent states of matter and designing devices that leverage nanoscale orbital currents for macroscopic functionalities, including tunable inductive elements, non-volatile memory, and low-power electronic components.

The COC circulating along the edges of $MnTe_6$ octahedra (**Fig. 1a**) provide a compelling framework for understanding the extraordinary phenomena observed in $Mn_3Si_2Te_6$ (**Fig.1b**) [1, 6-16]. These COC are stabilized by the SOI, as evidenced by a large magnetic anisotropy field of 13 T [6], and exhibit robust magnetoelastic coupling, leading to pronounced magnetostriction effects [1]. The SOI and magnetoelastic interactions underpin the unique COC physics in this material. Remarkably, even modest applied currents, *I*, strongly interact with the COC, giving rise to a series of highly unconventional, current-sensitive phenomena. These include a colossal magnetoresistance (CMR) exceeding $10^8$ in the absence of net magnetic polarization [1, 6, 8, 9], a current-sensitive Hall effect [7], anomalous non-linear I-V characteristics (**Fig. 1c**), and a striking first-order bistable switching between the COC and normal states (**Fig. 1d**) [1]. Each of these phenomena independently challenges conventional theoretical frameworks; their simultaneous manifestation points to the emergence of a unique, collective state of matter in $Mn_3Si_2Te_6$. Moreover, growing evidence suggests that similar chiral orbital orders may also exist in other quantum materials [11, 18-25], highlighting the broader relevance of this newly uncovered physics.

This study is primarily motivated by the observation of first-order bistable switching between the COC and normal states, accompanied by up to a three-order-of-magnitude change in resistance and remarkably slow switching timescales, lasting seconds to minutes (**Fig. 1d**) [1]. Such slow dynamics suggest a strong magnetoelastic coupling between the COC and the induced orbital moments, $\mathbf{M}_{\mathbf{COC}}$ (**Fig. 1a**), with the massive lattice. This behavior contrasts sharply with the ultrafast picosecond ($10^{-12}$ s) or femtosecond ($10^{-15}$ s) switching times typically observed in electronic systems [26]. To explore this slow dynamical regime, we focus on transport and magnetic responses at low frequencies, *f*, on the order of 1 Hz, matching the observed switching timescales (**Fig.1d**). Additional data were collected from $Mn_3(Si_{1-x}Ge_x)_2Te_6$ samples with $x = 0$ and 0.07 [17], because our previous studies indicate that Ge doping expands the *ab* plane of the unit cell, thereby enhancing the chiral orbital currents (COC) [1] and the associated orbital moments [9]. This structural tuning allows us to test the universality and robustness of the emergent inductive response across related compositions.

We first examine the I-V characteristics $Mn_3(Si_{1-x}Ge_x)_2Te_6$, which reveal key signatures of this novel inductive behavior. The I-V curves display two successive onsets of sharp S-shaped negative differential resistance (NDR), labeled $V_{NDR1}$ and $V_{NDR2}$, following an initial Ohmic regime in the absence of H || *c* axis (**Fig. 1c)** [1]. $V_{NDR1}$ is highly sensitive to the field orientation: For H || *c* axis, $V_{NDR1}$ *vanishes rapidly* when $\mu_o H_{||c} > 5$ T, leading to a highly conductive COC state with vanishingly small $\Delta V/\Delta I$ (**Fig. 1c**), directly linked to the CMR [1, 6]. In contrast, applying H || *a*-axis enhances $V_{NDR1}$. On the other hand, $V_{NDR2}$ responds similarly to both H || *c*-axis and H || *a*-axis, though the effect is more pronounced along the *c*-axis (**Fig. 1c**). Comparable behavior to $V_{NDR2}$ is also seen in other materials with strong spin-orbit interactions, such as $Sr_2IrO_4$ [27]. Importantly, only $V_{NDR1}$ is closely tied to the presence of the COC, making it central to the discussion that follows.

The I-V characteristics exhibit a remarkable sensitivity to both the temporal and frequency, *f*, profiles of the applied current. We therefore focus on representative I-V data at selected frequencies and magnetic fields along the *c*-axis [17]. For frequencies *f* between 0 and 12 Hz, strong hysteresis loops are consistently observed in the I-V curves, indicating a clear lag between the voltage and current responses. Typically, increasing *I* leads to higher *V*, and subsequently reducing *I* lowers *V*, resulting in a standard "counterclockwise" hysteresis loop. This behavior is common in magnetic materials and electronic systems [e.g., 27].

Strikingly, under a magnetic field applied along the *c*-axis, the hysteresis loop reverses direction, producing an unusual "clockwise" loop in the I-V characteristics. In this regime, increasing *I* produces little change in *V*, while decreasing *I* leads to a dramatic increase in *V* by as much as three orders of magnitude. It is important to note that this reversed hysteresis emerges only when H is applied along the *c*-axis, consistent with the field orientation required to stabilize the COC state [1]. Furthermore, at H || *c*-axis, $Mn_3Si_2Te_6$ becomes highly conductive, effectively ruling out capacitive effects as an explanation for the observed behavior [17, SFig. 1].

We emphasize that these results including nonlinear and hysteretic responses are not caused by Joule heating. A comprehensive discussion, including sample-mounted temperature measurements, field orientation dependence, and power scaling is presented in the Supplemental Material [17].

**Figs. 2-4** compare the hysteresis loop reversal across three distinct scenarios: **A**, **B**, and **C** to establish the central finding: Conventional counterclockwise hysteresis loops are characteristic of the normal state, while the emergent clockwise loops are a hallmark of the coherent COC state responsible for the novel inductive response. To further elucidate the distinct inductive behavior associated with the COC state, we systematically compare the hysteresis loop reversal under three key experimental conditions:

**A. $\mu_o H_{||c} = 0$ (normal state) vs. $\mu_o H_{||c} \neq 0$ (COC state). Fig. 2** illustrates the I-V characteristics for $x = 0$ at $f = 0$ and 6.1 Hz at $\mu_o H_{||c} = 0$ and $\mu_o H_{||c} = 14$ T. At $\mu_o H_{||c} = 0$ (blue curves), no significant hysteresis appears at $f = 0$ (**Figs. 2a–2b**), but clear counterclockwise hysteresis loops develop at $f = 6.1$ Hz (**Figs. 2c-2d**), featuring distinct $V_{NDR1}$ and $V_{NDR2}$. However, at $\mu_o H_{||c} = 14$ T (red curves), $V_{NDR1}$ vanishes, and the I-V curve becomes nearly vertical during the increasing current sweep, indicating vanishingly small $\Delta V/\Delta I$. Crucially, during the decreasing $I$ sweep, $V$ rises sharply, forming a reversed, clockwise hysteresis loop (**Fig. 2d**). This direct comparison establishes that the clockwise hysteresis loop is a unique signature of the COC-induced inductance under H || $c$-axis.

**B. T > $T_C$ (normal state) vs. T < $T_C$ at $\mu_o H_{||c} \neq 0$ (COC state). Fig. 3a** shows the temperature evolution of the I-V curves for $x = 0.07$ under $\mu_o H_{||c} = 14$ T at $f = 3$ Hz. At T = 120 K (> $T_C$), the system displays conventional counterclockwise hysteresis (also see SFig.2). However, for T = 2 K and 20 K (< $T_C$), the I-V curves exhibit prominent clockwise hysteresis loops, consistent with the emergence of the COC state. The corresponding resistance $R$, extracted from the I-V data, shows a nearly three-order-of-magnitude increase during the decreasing $I$ sweep compared to the increasing sweep (**Fig. 3b**). This behavior directly reflects the inductive response driven by the rigidity of the coherent COC domains.

**C. H || a vs. H || c.** Finally, **Fig. 3c** compares the I-V curves for $x = 0$ at $f = 1$ Hz and T = 20 K under $\mu_o H = 5$ T applied along the $c$-axis (red curve) and $a$-axis (blue curve). With H || $a$-axis, the system exhibits a normal counterclockwise hysteresis loop with distinct $V_{NDR1}$ and $V_{NDR2}$, resembling behavior observed at zero field. In contrast, applying H || $c$-axis induces a clockwise hysteresis loop, further confirming that the COC state and the associated inductive behavior requires a $c$-axis magnetic field.

These three scenarios firmly establish the link between the emergence of coherent COC domains and the novel inductive response, characterized by reversed hysteresis behavior and a dramatic increase in V upon decreasing *I*. This new form of inductance fundamentally differs from conventional geometric or kinetic inductance and highlights the critical roles of SOI, magnetoelastic coupling, and field orientation in stabilizing the COC state.

The unique I-V characteristics and hysteresis behavior observed under varying current and magnetic field conditions can be quantitatively modeled using a simple circuit analogy that incorporates both resistive and inductive elements in series. This model, based on the experimental data for $x = 0$ at 6.1 Hz, 14 T, and 10 K (**Fig. 2d**, red curve), successfully reproduces the key features of the observed inductive response: The inductive response is negligible during the increasing current sweep but becomes highly pronounced when the current is reduced, inducing a substantial electromotive force, *emf.* As shown in **Fig. 3d**, the calculated inductance remains effectively zero as *I* increases, consistent with the vertical I-V curves and vanishing $\Delta V/\Delta I$ in the COC state. However, during the decreasing current sweep, the model reveals a significant inductive contribution, corresponding to an effective inductance on the order of millihenries - several orders of magnitude larger than the electromagnetic inductance values reported for helical-spin magnets or related devices in recent studies [2, 28, 29]. This modeling outcome aligns closely with the experimental observations, reinforcing the interpretation that coherent COC domains act as mesoscopic inductive coils resisting rapid changes in current.

A closer analysis of the I-V data reveals the existence of a threshold current, $I_{Ind}$, marking the onset of the inductive response. For $I < I_{Ind}$, the system remains highly conductive with reversible I-V behavior and no apparent hysteresis. When $I > I_{Ind}$, the system enters a regime characterized by clockwise hysteresis and significant inductance during current reduction (**Fig. 4a-4c**).

By systematically mapping $I_{Ind}$ as a function of $f$, we construct a phase diagram for the COC-induced inductance (**Fig. 4d**). This diagram clearly shows that a minimum threshold current is required to reconfigure the COC domains and trigger the inductive state. Moreover, the inductive response vanishes for $f > 12$ Hz (SFig. 3), consistent with the long switching timescales identified in the bistable behavior (**Fig. 1d**) [1], which is a defining feature of the slow dynamics inherent to the COC phase. Remarkably, despite the disappearance of inductance at high frequencies, the high-conductivity state associated with small $\Delta V/\Delta I$ persists well beyond 12 Hz, underscoring the stability of the trained COC domains against dynamic perturbations (SFig.3). Notably, while DC currents above ~ 4 mA destroy the COC state, similar AC amplitudes do not, since the AC current is not sustained at high values but rather cycles through them. Instead, rising AC current reconfigures the COC into coherent domains, and their rigidity resists reversal of the first-order transition during current reduction, producing the large *emf.*

Building on this framework, the evolution of the COC state underpins the observed unconventional inductive behavior. This evolution proceeds through three distinct stages, each defined by specific experimental conditions and a characteristic response of the COC domains. These stages clarify how coherent COC structures form and generate the large induced *emf* observed during current reduction, highlighting the critical roles of current direction, magnetic field orientation, and strong magnetoelastic coupling (**Fig. 5a**). These stages are as follows:

**Stage I: Random COC, No Net Inductance ($\mu_o H_{||c} = 0$).** Under ambient conditions, COC domains are randomly oriented, producing no net inductance [1] (**Fig. 5a,** Stage I). Applying $I$ initiates partial COC reorganization, marked by the onset of $V_{NDR1}$ and a reduction in resistance $R$ (**Fig. 5a**). It is important to note that increasing $I$ ($< I_C$) significantly reduces the resistivity ρ without H [1].

**Stage II: Formation of Coherent COC "Coils" (Increasing $I$ + $\mu_o H_{||c} \neq 0$)**. With H || $c$-axis and increasing $I$, COC domains reorganize through first-order transitions near $V_{NDR1}$, expanding domains circulating in one direction while suppressing the opposite (**Fig. 5a,** Stage II). This process trains the COC into extended, coherent structures, strengthening the $c$-axis $\mathbf{M_{COC}}$ and enhancing magnetoelastic coupling, effectively forming mesoscopic inductive "coils."

**Stage III: Inductive emf from COC Coils (Decreasing $I$ + $\mu_o H_{||c} \neq 0$)**. Reducing $I$ attempts to reverse the trained COC state, but the reversal of first-order transitions is prohibited. Strong magnetoelastic coupling and inertia prevent this change, leading the coherent COC to generate a substantial *emf* opposing the current decrease (**Fig. 5a**, Stage III). This accounts for the sharp rise in voltage $V$ and resistance $R$ as $I$ decreases.

The presence of these COC "coils" also explains the pronounced diamagnetic responses observed in AC susceptibility $\chi'$ at low temperatures and $f$ = 10 Hz (**Fig. 5b**). These responses strengthen at $\mu_o H_{||c}$ > 11 T but vanish at $f$ = 10 kHz (**Fig. 5c**), consistent with the low-frequency dynamics of the COC. Such strong diamagnetism is highly unusual in metallic, magnetic systems.

The inductive response emerges only between a threshold current $I_{Ind}$ and the critical current $I_C$ . Below $I_{Ind}$ the COC remain incoherent; above $I_C$, the COC state collapses. This defines a finite window for the COC-induced inductance.

In summary, we reveal a previously unknown inductance mechanism in bulk $Mn_3Si_2Te_6$, enabled by the interplay of spin-orbit and magnetoelastic couplings that rigidly bind COC to the lattice. These coherent COC act as mesoscopic inductive elements, producing pronounced low-frequency electromagnetic responses and unexpected diamagnetic behavior. The slow reconfiguration dynamics, occurring over seconds to minutes, are not suited for high-speed switching, but are well-matched to emerging applications such as nonvolatile memory, neuromorphic computing, and adaptive circuit elements, where stability, tunability, and retention are prioritized. This discovery in a bulk material, without nanoscale engineering, challenges conventional electromagnetic models and offers new opportunities for developing emergent quantum devices.

**Notes on Joule Heating (**See Supplemental Material attached to the end of this paper**):**

A recent PRL *Letter* (https://doi.org/10.1103/ry2d-bgy; https://arxiv.org/abs/2502.11048) attributes the nonlinear transport and bistable switching we reported in $Mn_3Si_2Te_6$ [1] to Joule heating. This conclusion is incompatible with our previously published data and this work [1, 6,7, 9].

Most importantly, the *Letter's* measurements were performed at current densities on the order of **$10^3$–$10^4$ A/cm²**, where significant Joule heating is expected. In contrast, our measurements were conducted at **~1 A/cm²** [1, 6, 7, 9], several orders of magnitude lower, where direct thermometry confirms negligible temperature rise (< 5 K) [7]. Thermal effects that may occur at extreme current densities cannot be used to interpret our results obtained in this much lower-current regime, underscoring the need for a distinct interpretation of our results.

Joule heating is a gradual, isotropic process, independent of current direction or field orientation. In contrast, our experiments reveal abrupt, first-order transitions that are highly nonlinear, anisotropic, and field- and current-direction dependent – clear signatures that cannot be explained by simple thermal activation.

Because phenomena arising from chiral orbital currents (COC) represent a rapidly developing and impactful research area, it is important to clarify this distinction so that future discussions of current-induced transport in $Mn_3Si_2Te_6$ are guided by a mechanism consistent with the full body of experimental evidence. **Please see Supplemental Material attached to the end**

**of this paper.**

**ACKNOWLEDGEMENTS**

G.C. thanks Longji Cui, Minhyea Lee and Dan Dessau for useful discussions. This work is supported by U.S. National Science Foundation via Grant No. DMR 2204811. I.K. acknowledges support by U.S. Department of Energy Office of Science via Early Career Award DE-SC0025478.

**DATA AVAILABILITY**

The data that support the findings of this Letter are not publicly available upon publication because it is not technically feasible and/or the cost of preparing, depositing, and hosting the data would be prohibitive within the terms of this research project. The data are available from the corresponding author upon reasonable request.

**FIGURE LENGENDS**

**Fig.1. Underlying structural and physical properties for x = 0. a,** The crystal and magnetic structure of $Mn_3Si_2Te_6$ [1]. The colored circles and vertical arrows indicate the *ab*-plane COC and induced $\mathbf{M}_{COC}$, respectively; different colors indicate different magnitudes of the *ab-plane* COC and $\mathbf{M}_{COC}$; the green triangles denote *off-ab-plane* COC that are insignificant [1]; the faint cylindrical arrows are Mn spins [3]. **b,** A representative single crystal of $Mn_3Si_2Te_6$. **c**, The I-V curves driven by DC at H = 0 T (black), $\mu_oH_{||c}$ = 14 T(red) and $\mu_oH_{||a}$ = 14 T (blue). **d**, Time-dependent bistable switching: The *a*-axis voltage V as a function of time t at T = 10 K for $\mu_oH_{||c}$ = 7 T [1]. Note the long switching time of ~ 120 s.

**Fig. 2. DC vs. AC Currents for I-V characteristic at $\mu_o H_{||c}$ = 0, 14 T, and T = 10 K for x = 0. a-d,** The I-V curves at $f$ = 0 (a, b) [1] and 6.1 Hz (c, d). Note that **b** and **d** show the first quadrant for clarity and that the change of the I-V hysteresis loops from counterclockwise at $\mu_o H_{||c}$ = 0 (blue) to clockwise at $\mu_o H_{||c}$ = 14 T (red) (**c-d**). No significant hysteresis for $f$ = 0 (**a-b**).

**Fig. 3. The temperature and field dependence of inductance - The I-V characteristic for T > $T_C$ and T < $T_C$ at $\mu_o H_{||c}$ = 14 T, and for H || *a* axis and H || *c* axis; the outcome of the modeling. a,** The I-V curves for x =0.07 at $f$ = 3 Hz and $\mu_o H_{||c}$ = 14 T for selected T = 2, 20 and 120 K. Note the change of the I-V hysteresis loops from counterclockwise at T > $T_C$ to clockwise at T < $T_C$. **b,** The resistance R converted from the I-V curves at 5 K (the I-V curve at 5 K is not shown). Note *R* increases due to decreasing *I*. **c,** The I-V curves for x = 0 at $f$ = 1.0 Hz and 20 K for $\mu_o H_{||c}$ = 5 T (red) and $\mu_o H_{||a}$ = 5 T (blue). **d,** Selected result of the electronic modeling based on the experimental data at $f$ = 6.1 Hz and $\mu_o H_{||c}$ = 14 T (Fig.2d, red curve). Note the inductance is indiscernible when *I* is rising, but significantly large, on the order of mH, when *I* is reducing.

**Fig. 4. The threshold behavior and dynamic phase diagram of inductance. a-c,** Selected I-V curves for x =0.07 at $f$ = 0.3, 3.0 and 6.1 Hz that reveals a threshold current $I_{ind}$ for induction. **d**, The phase diagram of $I_{Ind}$ as a function of $f$ for H || *c* axis. Note that at $I < I_{Ind}$, small $\Delta V/\Delta I$ retains, regardless of increasing or decreasing *I* and that the induction vanishes at $f$ >12 Hz, but small $\Delta V/\Delta I$ extends well beyond 12 Hz.

**Fig. 5. Schematics of reconfigurations of the COC and induction. a,** The schematics illustrating Stage I: $\mu_o H_{||c}$ = 0, Stage II: Constant $\mu_o H_{||c} \neq 0$ and rising *I*, and Stage III: Constant $\mu_o H_{||c} \neq 0$ and reducing *I*. The coil in Stage III is imaginary. **b, c**, The real part of AC susceptibility $\chi$' for x = 0 at $f$ = 10 Hz (b) and 10kHz (c), and T = 0.15 and 0.5 K. Note the strong diamagnetic response and

the rapid drop in $\chi$' at $\mu_o H_{||c} > 11$ T for $f = 10$ Hz. This behavior weakens or vanishes for $f = 10$ kHz.

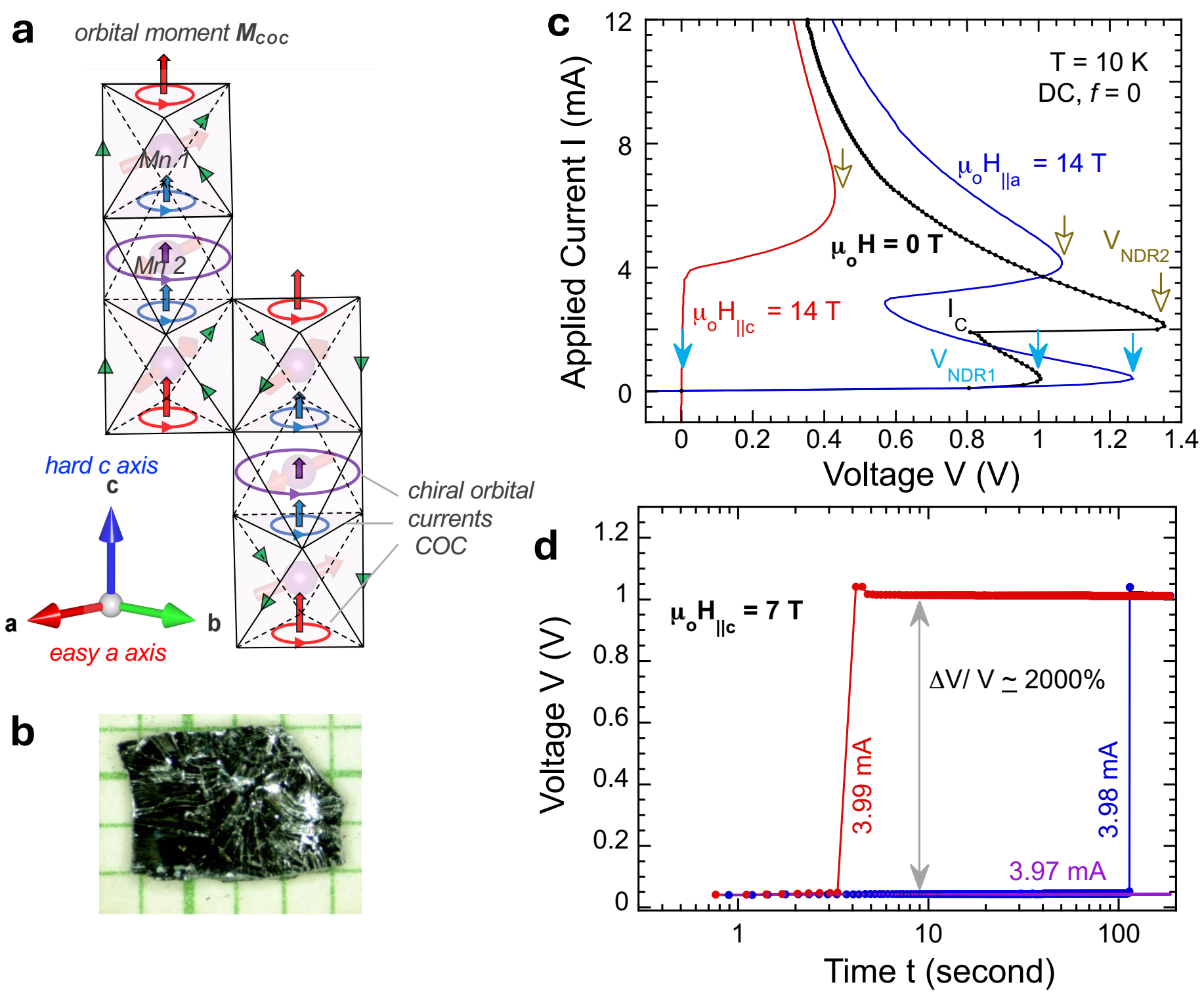
a
orbital moment M_COC
Mn 1
Mn 2
hard c axis
c
a
b
easy a axis
chiral orbital
currents
COC
b
c
T = 10 K
DC, f = 0
μoH||a = 14 T
μoH = 0 T
V_NDR2
μoH||c = 14 T
I_C
V_NDR1
Applied Current I (mA)
Voltage V (V)
d
μoH||c = 7 T
ΔV/ V ≃ 2000%
3.99 mA
3.98 mA
3.97 mA
Voltage V (V)
Time t (second)

Figure 1

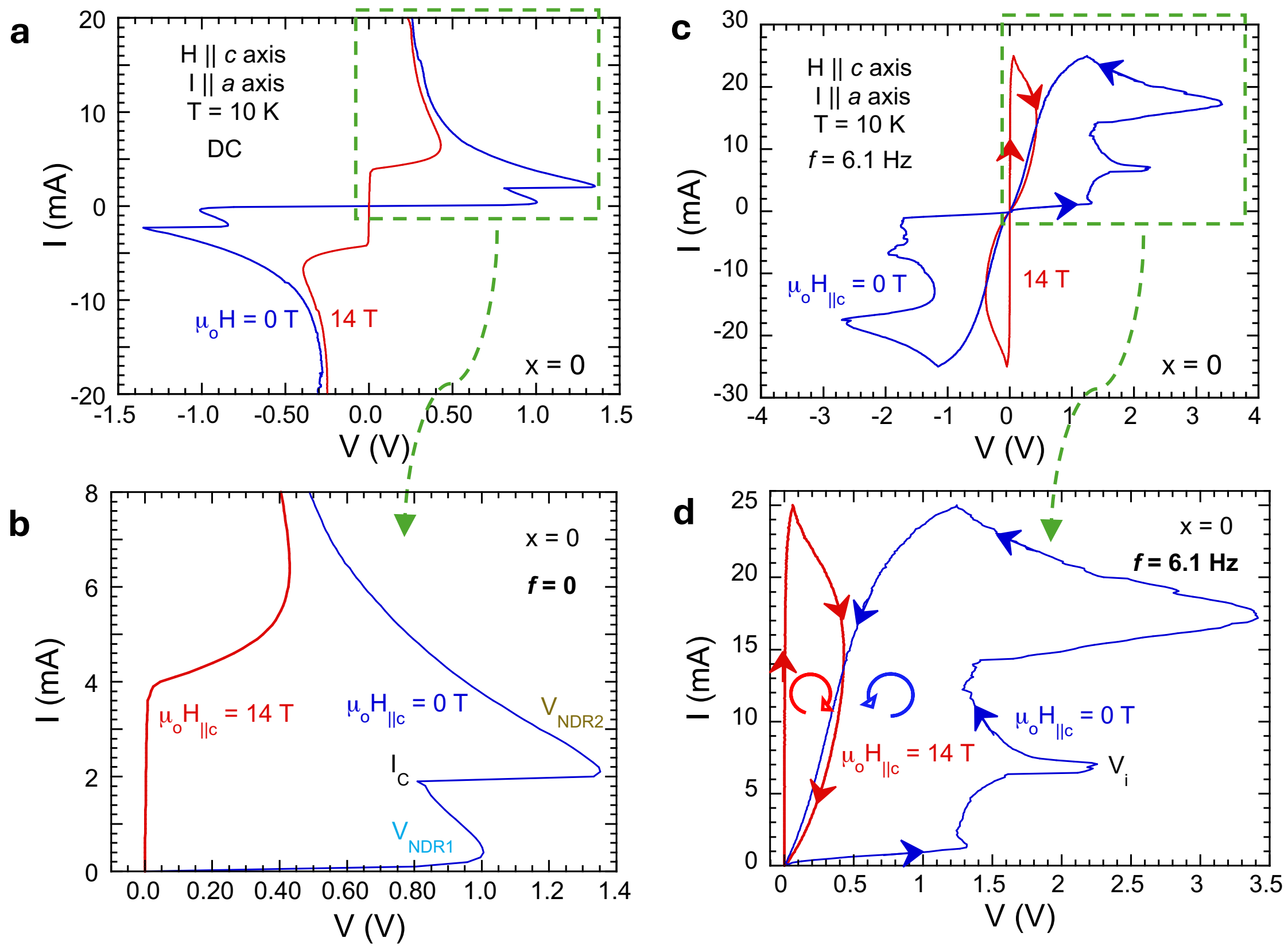

a
H || c axis
I || a axis
T = 10 K
DC
μoH = 0 T
14 T
x = 0
I (mA)
V (V)
b
x = 0
f = 0
μoH||c = 14 T
μoH||c = 0 T
VNDR2
IC
VNDR1
I (mA)
V (V)
c
H || c axis
I || a axis
T = 10 K
f = 6.1 Hz
μoH||c = 0 T
14 T
x = 0
I (mA)
V (V)
d
x = 0
f = 6.1 Hz
μoH||c = 0 T
μoH||c = 14 T
Vi
I (mA)
V (V)

Figure 2

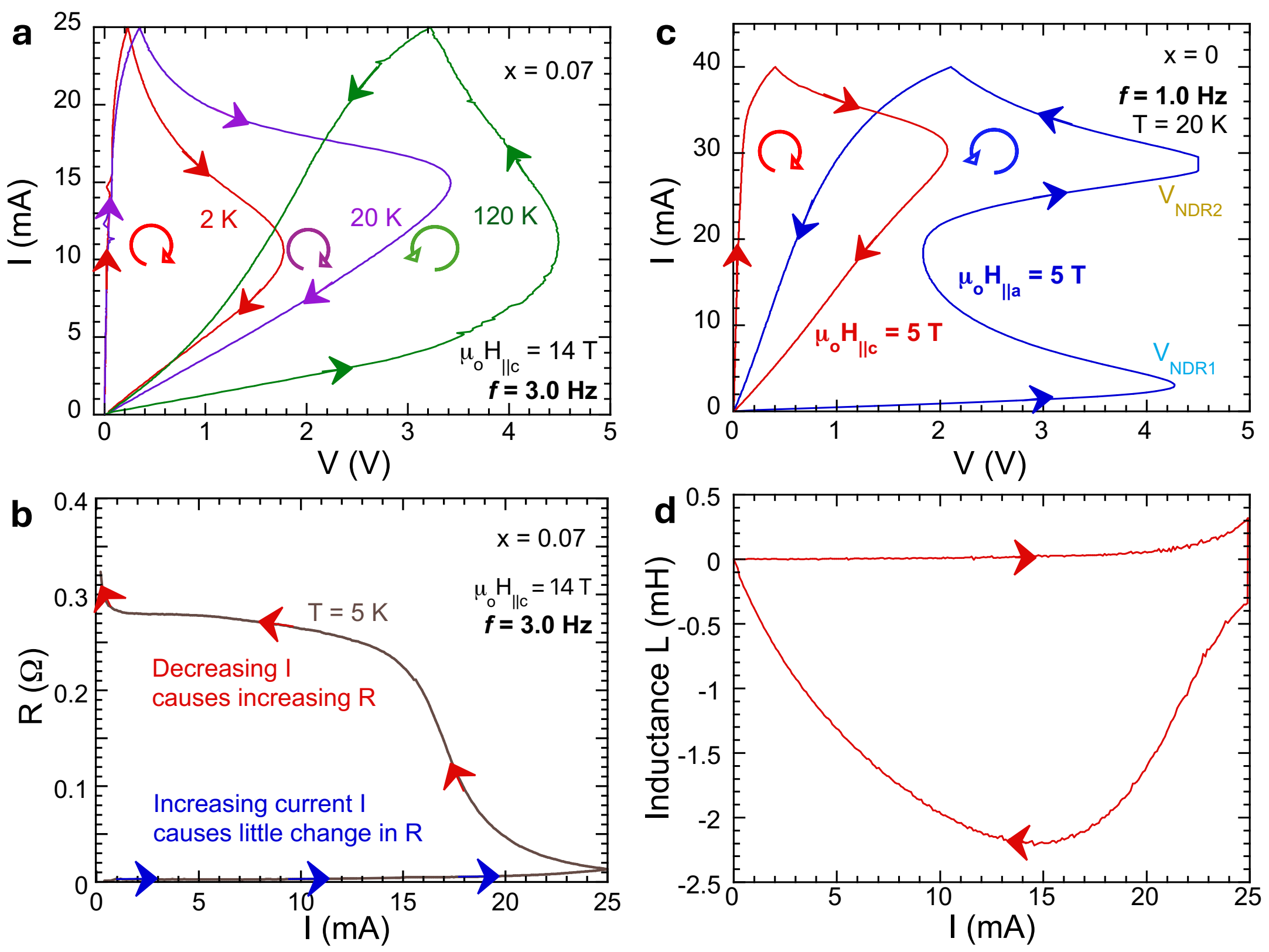

a
x = 0.07
2 K
20 K
120 K
μoH||c = 14 T
f = 3.0 Hz
I (mA)
V (V)
b
x = 0.07
T = 5 K
μoH||c = 14 T
f = 3.0 Hz
Decreasing I causes increasing R
Increasing current I causes little change in R
R (Ω)
I (mA)
c
x = 0
f = 1.0 Hz
T = 20 K
VNDR2
VNDR1
μoH||a = 5 T
μoH||c = 5 T
I (mA)
V (V)
d
Inductance L (mH)
I (mA)

Figure 3

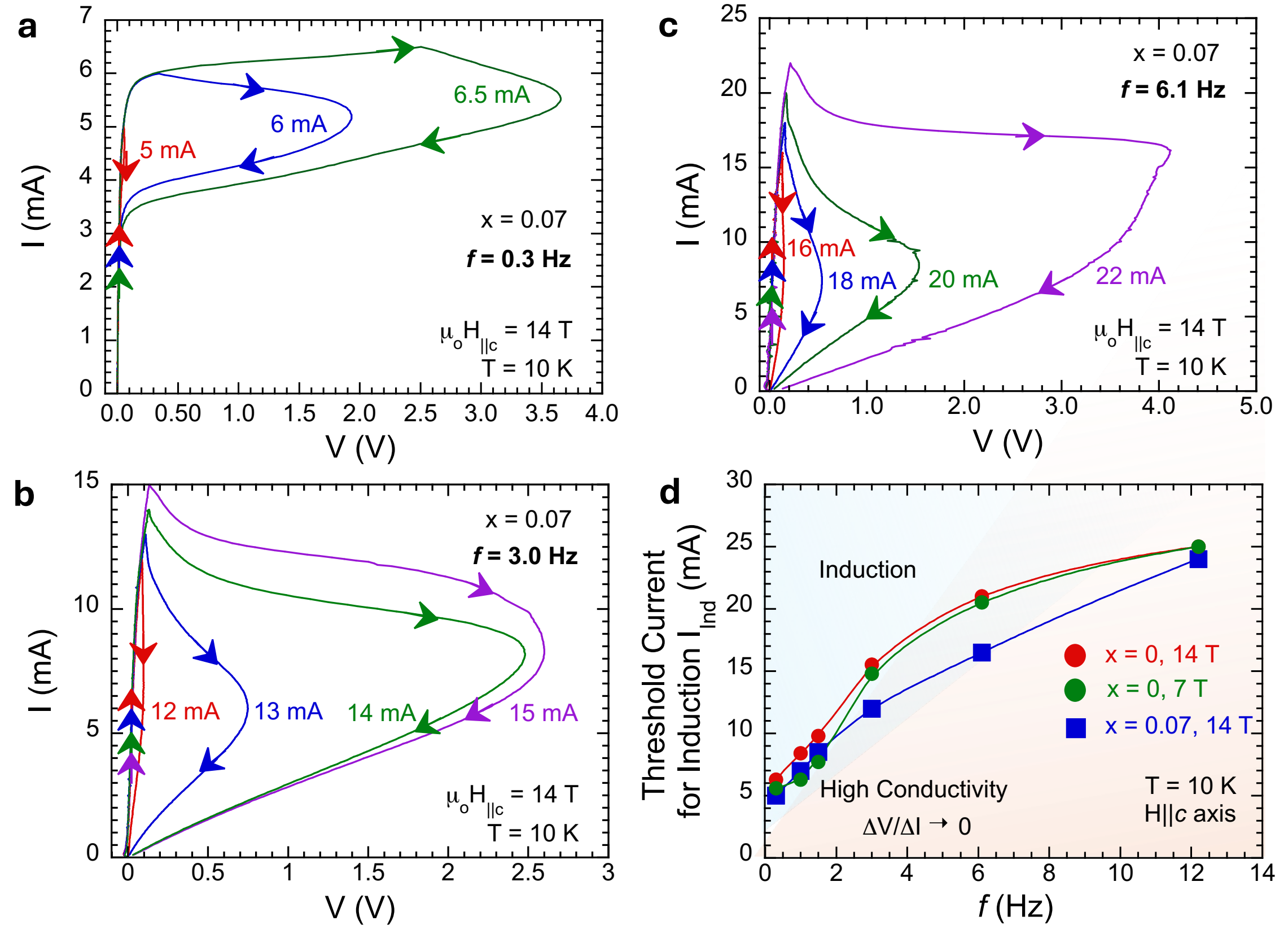

a
5 mA
6 mA
6.5 mA
x = 0.07
f = 0.3 Hz
μoH||c = 14 T
T = 10 K
I (mA)
V (V)
b
12 mA
13 mA
14 mA
15 mA
x = 0.07
f = 3.0 Hz
μoH||c = 14 T
T = 10 K
I (mA)
V (V)
c
16 mA
18 mA
20 mA
22 mA
x = 0.07
f = 6.1 Hz
μoH||c = 14 T
T = 10 K
I (mA)
V (V)
d
Induction
High Conductivity
ΔV/ΔI → 0
x = 0, 14 T
x = 0, 7 T
x = 0.07, 14 T
T = 10 K
H||c axis
Threshold Current for Induction I_Ind (mA)
f (Hz)

Figure 4

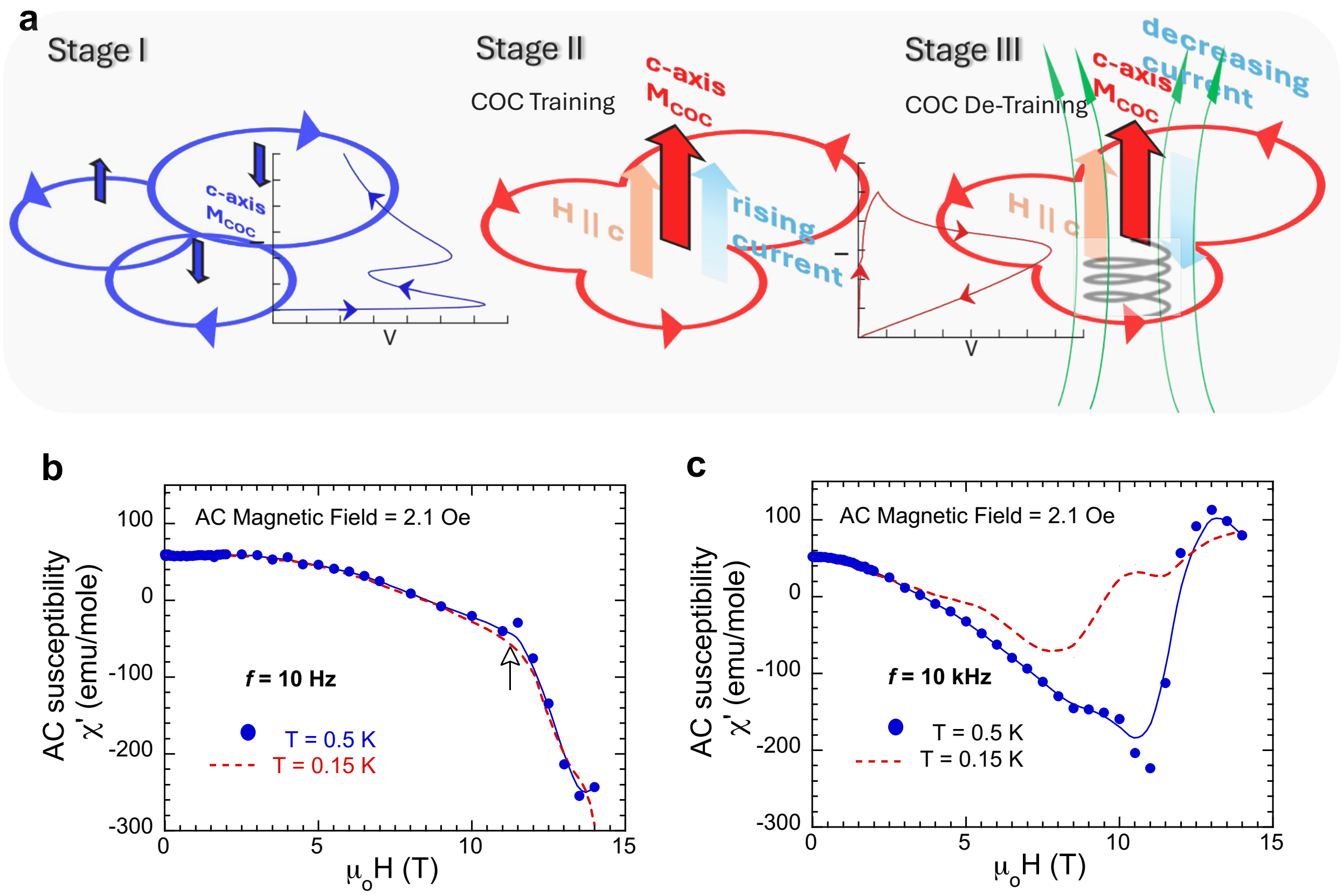
a
Stage I
c-axis
Mcoc
V
Stage II
COC Training
c-axis
Mcoc
H || c
rising
current
Stage III
COC De-Training
c-axis
Mcoc
decreasing
current
H || c
V
b
AC Magnetic Field = 2.1 Oe
f = 10 Hz
T = 0.5 K
T = 0.15 K
AC susceptibility
χ' (emu/mole)
μoH (T)
c
AC Magnetic Field = 2.1 Oe
f = 10 kHz
T = 0.5 K
T = 0.15 K
AC susceptibility
χ' (emu/mole)
μoH (T)

Figure 5

SUPPLEMENTAL MATERIAL

# Emergent Inductance from Chiral Orbital Currents in a Bulk Ferrimagnet

Gang Cao[1,2*], Hengdi Zhao[1], Yu Zhang[1], Alex Fix[1], Tristan R. Cao[1], Dhruva Ananth[1], Yifei Ni[1], Gabriel Schebel[1], Rahul Nandkishore[1,3], Itamar Kimchi[4], Hua Chen[5], Feng Ye[6] and Lance E. DeLong[7]

[1]*Department of Physics, University of Colorado at Boulder, Boulder, CO 80309*

[2]*Center for Experiments on Quantum Materials, University of Colorado at Boulder, Boulder, CO 80309*

[3]*Center for Theory of Quantum Matter, University of Colorado at Boulder, Boulder, CO 80309*

[4]*School of Physics, Georgia Institute of Technology, Atlanta, GA 30332, USA*

[5]*Department of Physics and Astronomy, Colorado State University, Fort Collins, CO 80521*

[6]*Neutron Scattering Division, Oak Ridge National Laboratory, Oak Ridge, Tennessee 37831*

[7]*Department of Physics and Astronomy, University of Kentucky, Lexington, KY 40506*

## I. Experimental Details

Single crystals of $Mn_3Si_2Te_6$ were grown using a flux method. Measurements of crystal structures were performed using a Bruker Quest ECO single-crystal diffractometer with an Oxford Cryosystem providing sample temperature environments ranging from 80 K to 400 K. Chemical analyses of the samples were performed using a combination of a Hitachi MT3030 Plus Scanning Electron Microscope and an Oxford Energy Dispersive X-Ray Spectroscopy (EDX). The measurements of the electric resistivity and I-V characteristic were carried out using a Quantum Design (QD) Dynacool PPMS system having a 14-Tesla magnet and a set of external Keithley meters that provides current source and measures voltage with a high precision.

It is important to point out that the I-V curves presented in the main text are current-driven. I-V curves at $f = 0$ Hz are taken using a pair of Keithley meters, and those at $f > 0$ Hz by the QD

Dynacool PPMS, whose electronics applies a triangular excitation waveform that always starts and ends at zero bias (see Manual for QD PPMS Electrical Transport Option).

In our experiments, the contact resistance was measured to be on the order of 10 Ω, at least 4 orders of magnitude smaller than the sample resistance in the insulating state (>$10^5$ W). Importantly, these measurements were made using standard four-probe configurations, with current and voltage leads separated to eliminate contact contributions to voltage measurements.

## II. Magnetic Structure

With trigonal symmetry (P-31c), $Mn_3Si_2Te_6$ ferrimagnetically orders at a transition temperature $T_C$ = 78 K with an easy *a* axis, and a hard *c* axis. It features a noncollinear magnetic structure with the magnetic space group *C2'/c'*, where the Mn spins lie predominantly within the *ab* plane with a 10°-tilting toward the *c* axis in ambient conditions. The noncollinear magnetic structure simultaneously breaks mirror and time reversal symmetries, which is essential for the COC to form below $T_C$. The COC circulate on the edges of $MnTe_6$ octahedra but predominantly within the *ab* plane and therefore generate the *c*-axis orbital moments, $\mathbf{M_{COC}}$.

## III. No Capacitive Effects

The capacitance, $C_p$, and the associated dissipation factor, DF, were measured using a QuanTech LCR meter. SFig.1a illustrates $C_p$ and DF as a function of magnetic field H || *c* axis at T = 3 K and frequency *f* = 2 MHz. The behavior of DF exhibits significant oscillations with changes in H drastically oscillates with H, which is atypical of a capacitor. DF is defined as DF = R/X where capacitive reactance X = 1/2p *f* $C_p$. Typically, DF is expected to increase with *f*. However, in this case, DF actually decreases with increasing *f*, as shown in SFig.1b. It is also noteworthy that DF diverges at low *f*, indicating that electrons are mobile and the charge polarization necessary for capacitor function is not being established. More generally, $Mn_3Si_2Te_6$ becomes highly conductive at H || c axis, which makes any capacitive effects impossible.

## IV. Joule Heating Plays No Role in the COC State

A recent PRL *Letter* (https://doi.org/10.1103/ry2d-bgy; https://arxiv.org/abs/2502.11048) attributes the nonlinear transport and bistable switching we reported in $Mn_3Si_2Te_6$ [1] to Joule heating. This conclusion is incompatible with our previously published data and new measurements [1-5].

Most importantly, the *Letter's* measurements were performed at current densities on the order of **$10^3$–$10^4$ A/cm²**, where significant Joule heating is expected. In contrast, our measurements were conducted at **~1 A/cm²** [1-5], several orders of magnitude lower, where direct thermometry confirms negligible temperature rise (< 5 K) [3]. Thermal effects that may occur at extreme current densities cannot be used to interpret our results obtained in this much lower-current regime, underscoring the need for a distinct interpretation of our results.

Joule heating is a gradual, isotropic process, independent of current direction or field orientation. In contrast, our experiments reveal abrupt, first-order transitions that are highly nonlinear, anisotropic, and field- and current-direction dependent – clear signatures that cannot be explained by simple thermal activation.

Because phenomena arising from chiral orbital currents (COC) represent a rapidly developing and impactful research area, it is important to clarify this distinction so that future discussions of current-induced transport in $Mn_3Si_2Te_6$ are guided by a mechanism consistent with the full body of experimental evidence.

In the following, we present multiple, independent lines of evidence, including pulsed measurements, anisotropic field response, power scaling, differential resistance measurements and direct temperature measurements with a sample-mounted Cernox sensor, to demonstrate that the novel phenomena arise from intrinsic COC, conclusively eliminating Joule heating as their origin [1-5].

**Evidence Against Joule Heating**

1. **Bistable Switching and Field Orientation** [1]

   In **SFig. 4**, tiny current steps (ΔI ≈ 0.005-0.01 mA) trigger discrete voltage jumps (ΔV up to 0.99 V) that depend strongly on field orientation: Robust for H || *c* (**SFig.4b**) but absent for H || *a*, where Ohmic behavior returns (**SFig.4c**). ***If this switching were due to Joule heating, V or R would decrease rather than increase because R at H = 0 follows an insulating behavior*** (***i.e., the higher temperature corresponds to lower resistance***). Moreover, the switching is between two values of V, thus the bistable switching essentially independent of applied currents [1]. Joule heating, being isotropic and continuous, cannot produce such switching.

2. **Nonlinearity in Differential Resistance**

   In **SFig. 5**, differential resistance dV/dI vs DC current up to 30 mA at $f$ = 1 Hz with a 15 mA AC modulation shows pronounced, symmetric nonlinearity about I = 0 and sharp features near well-defined threshold currents, which are clear signatures of intrinsic, current-driven reconfiguration of orbital textures. Importantly, no monotonic resistance drop or runaway behavior is observed at high bias, ruling out Joule-heating artifacts. The response is fully reversible and frequency dependent. In particular, the strongly nonlinear yet symmetric dV/dI curves are incompatible with heating, which would produce a smooth, monotonic resistance decrease with increasing current.

3. **Power Inversely Proportional to Current – Current-Direction-Dependent Resistance** [5]

In **SFig. 6**, the resistance R remains essentially unchanged and small (~ 0.001 Ohm) with increasing current I but rises rapidly by nearly three-orders of magnitude ***only when current is reduced***, and power dissipation increases as current decreases: e.g., 2.0 mW at I = 20 mA vs. 4.3 mW at I = 4 mA. This is clearly inconsistent with Joule heating: Joule heating, which scales with $I^2R$, would rise when the current I increases, and fall when the current I decreases.

4. **Extreme Field Anisotropy** [1,5]

In **SFig. 7** I-V curves at H || *a* and H || *c* should behave similarly, independent of the field orientation if Joule heating were a driving force. But the I-V curves in **SFig.7** respond vastly differently to different field orientations (see red curve for H || *c* and blue curve for H || *a*). At I = 2 mA, power differs by three orders of magnitude between H || *c* and H || *a* (0.008 mW vs. 2.6 mW), which cannot arise from an isotropic heating mechanism.

5. **Hysteresis Loop Reversal of I-V Curves** [5]

Similarly, in **SFig. 8**, the I-V curves at H || *a* and H || *c* respond entirely differently to different field orientations: H || *c* causes a ***clockwise loop*** indicating the emergent inductive behavior [5], i.e., increasing current I only slight changes voltage V, whereas decreasing current I induces an increase in V. In contrast, H|| *a* results in a ***counterclockwise loop***: increasing I causes two negative differential resistance transitions while decreasing I leads to a decrease in V (see red curve for H || *c* and blue curve for H || *a*). The contrasting behaviors for H || *a* and H || *c* are once again inconsistent with Joule heating.

6. **Pulsed vs. Continuous Current Measurements** [5]

In **SFig. 9**, I-V curves under pulsed (0.1 s on/off) and DC current are nearly identical up to 12 mA, confirming that thermal accumulation plays no role.

7. **Direct Thermometry** [3]

Cernox-based measurements with a **Cernox thermometer attached to a single-crystal sample** (**SFig.10**) show the increase of temperature $\Delta T < 5$ K even at the highest currents used and decreasing $\Delta T$ with increasing field, incompatible with thermal runaway (**SFigs. 11-12**). Note that ***the current density is on the order of 1 A/cm²*** [3].

**Conclusion**

Our comprehensive data, spanning anisotropic field dependence, first-order switching, inverse power scaling, hysteretic I-V loops, pulsed/DC equivalence, and direct thermometry, conclusively rule out Joule heating as the mechanism for the nonlinear transport in $Mn_3Si_2Te_6$ [1-5]. These effects originate from reconfigurable COC domains, representing a new type of collective electronic state [5].

Because COC phenomena are a rapidly developing frontier, establishing the correct microscopic mechanism is essential for informing future experimental and theoretical efforts. We offer this clarification with the aim of fostering clearer distinctions between intrinsic chiral-orbital-current phenomena and thermal artifacts.

**Figure Captions**

**SFig. 1. Capacitance $C_p$ and dissipation factor DF for x = 0. a,** $C_p$ and DF at T = 3 K and *f* = 2 MHz as a function of magnetic field H aligned along the c axis. b, DF at T = 10 K and H = 0 as a function of *f*. Note that DF diverges as *f* approaches zero, suggesting the absence of charge polarization needed for a capacitor.

**SFig. 2. Counterclockwise hysteresis loop above $T_C$ at both $\mu_o H_{||c}$ = 0 and 14 T for x = 0.07.** The hysteresis loop is *counterclockwise* at both $\mu_o H_{||c}$ = 0 and 14 T, indicating the disappearance of the inductance.

**SFig. 3**. **The I-V characteristic at $\mu_o H_{||c}$ = 0, 14 T, and T = 10 K for x = 0.** At *f* = 12.2 Hz, the hysteresis loop almost vanishes. But the high conductive state is maintained.

**SFig. 4. Time-dependent bistable switching [1]:** The *a*-axis voltage $V_a$ as a function of time t at T = 10 K for **(a)** H = 0, **(b)** $\mu_o H_{||c} = 7$ T and **(c)** $\mu_o H_{||a} = 7$ T [1]. Note that there are only two values of V independent of applied currents I.

**SFig.5. Nonlinearity in Differential Resistance:** dV/dI vs DC current up to 30 mA at $f = 1$ Hz with 15mA AC modulation shows pronounced, symmetric nonlinearity about I=0I=0 with sharp features near threshold currents, signatures of intrinsic current-driven reconfiguration of orbital textures. Crucially, no monotonic resistance decrease or runaway behavior is observed at high bias, ruling out Joule-heating artifacts. The nonlinearity is reversible and frequency dependent, consistent with electronic origin rather than thermal effects.

**SFig. 6. Resistance R rises with reducing current I [5]:** R remains essentially unchanged with increasing current but rises rapidly by nearly three-orders of magnitude only when current is reduced at $\mu_o H_{||c} = 14$ T and 5 K. Note that resistance R depends on current direction.

**SFig. 7. Highly anisotropic I-V curves [1,5]:** The I-V curves driven by DC at H = 0 T (black), $\mu_o H_{||c} = 14$ T(red) and $\mu_o H_{||a} = 14$ T (blue). Note the high sensitivity of I-V curves to field orientation [1,5].

**SFig. 8. Highly anisotropic I-V curves [5]:** The I-V curves at $f = 1.0$ Hz and 20 K for $\mu_o H_{||c} = 5$ T (red) and $\mu_o H_{||a} = 5$ T (blue). Note that the change of the I-V loop orientation due to the change of field orientation [5].

**SFig. 9. Comparison of I-V curves measured under continuous DC (black) and pulsed (red) current excitation at 0 T [5].** Pulsed data were collected using a 0.1 s on / 0.1 s off duty cycle. The nearly overlap of the curves up to 12 mA demonstrates that Joule heating is negligible under the measurement conditions.

**SFig. 10. Direct Temperature Measurements with a Sample-Mounted Cernox Sensor [3]:** A ***Cernox thermometer*** (gold, front) is ***thermally contacted with a single-crystal sample*** of $Mn_3Si_2Te_6$ (black, behind the Cernox). The thin gold wires are electrical leads electrically attached to the sample and Cernox with an EPO-TEK H20E epoxy (silver). The light brown background is GE varnish used to thermally anchor the Cernox and the sample. Note that H||c-axis and I || a-axis.

**SFig. 11. Temperature change ΔT (K) as functions of applied current I (mA) and current density J (A/cm$^2$)** at T = 30 K for representative magnetic fields $\mu_o H$ || *c* axis [3].

**SFig. 12. Temperature change ΔT (K) as functions of applied current I (mA) and current density J (A/cm$^2$)** at $\mu_o H_{||c}$ = 3 T for representative temperatures T [3].

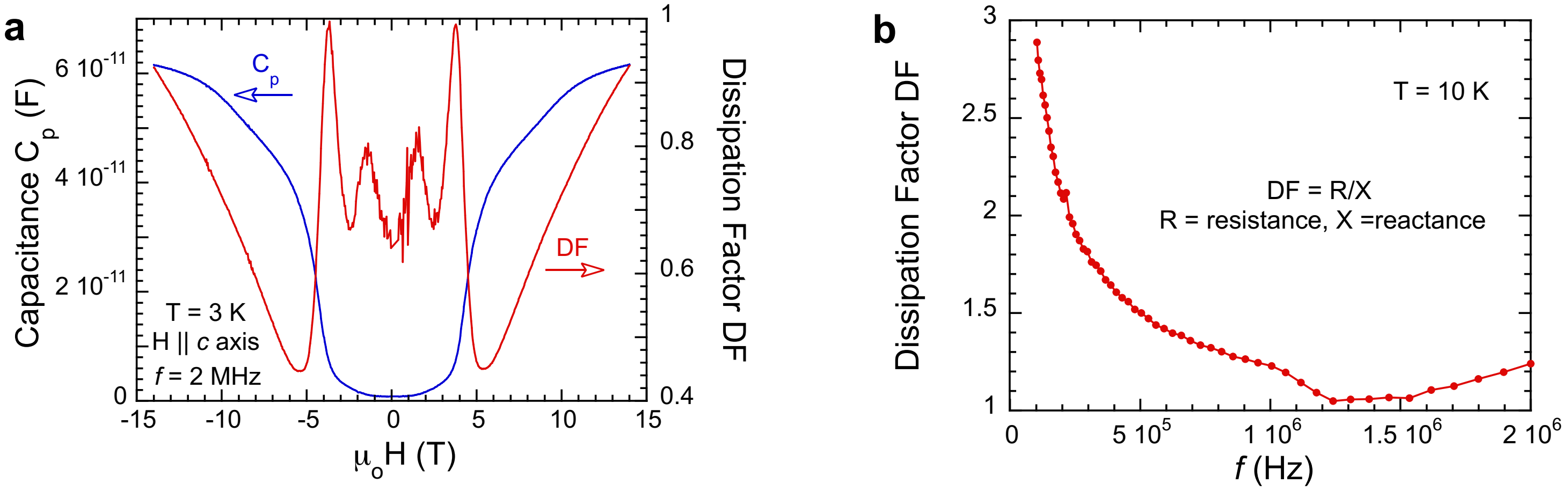

SFigure 1

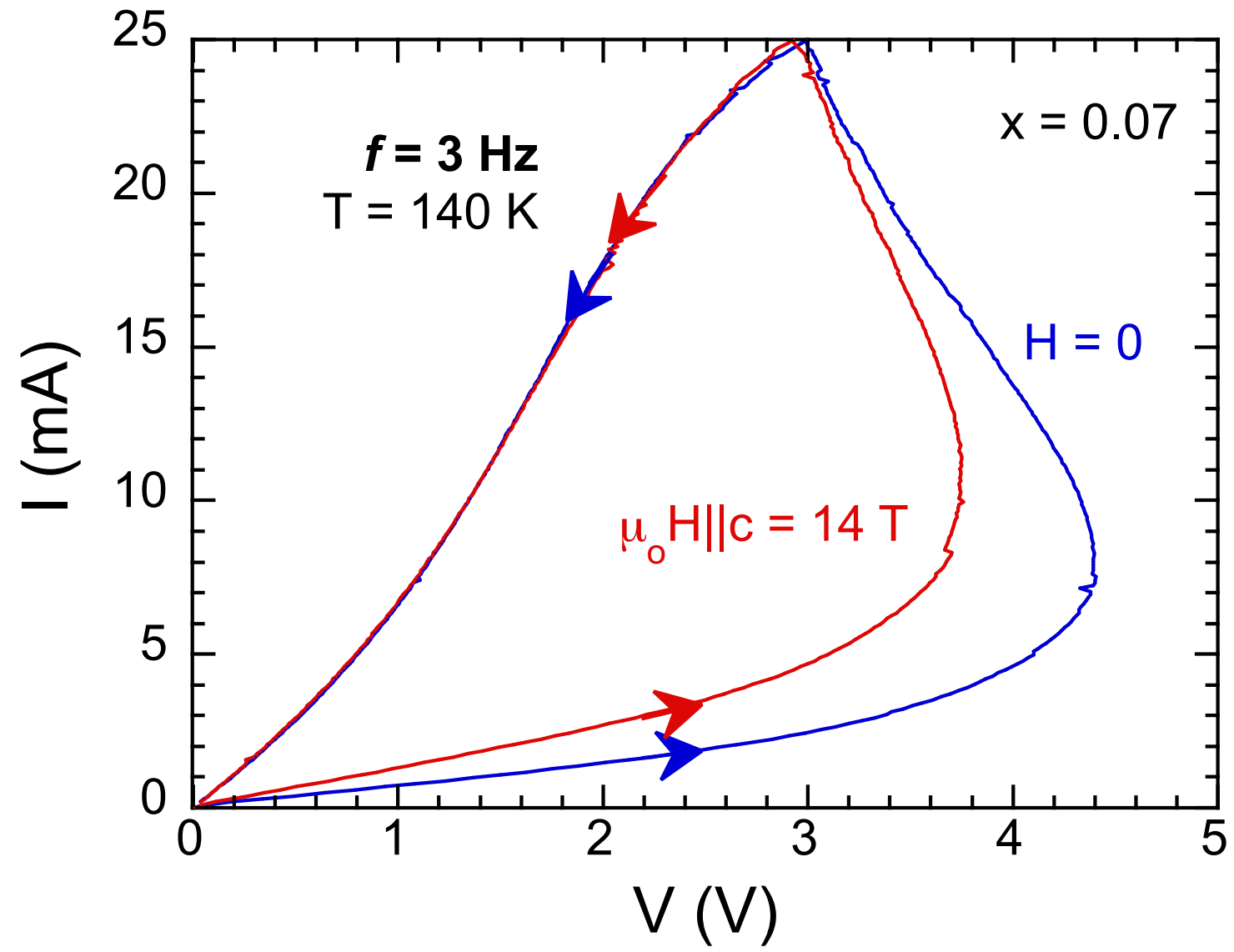

SFigure 2

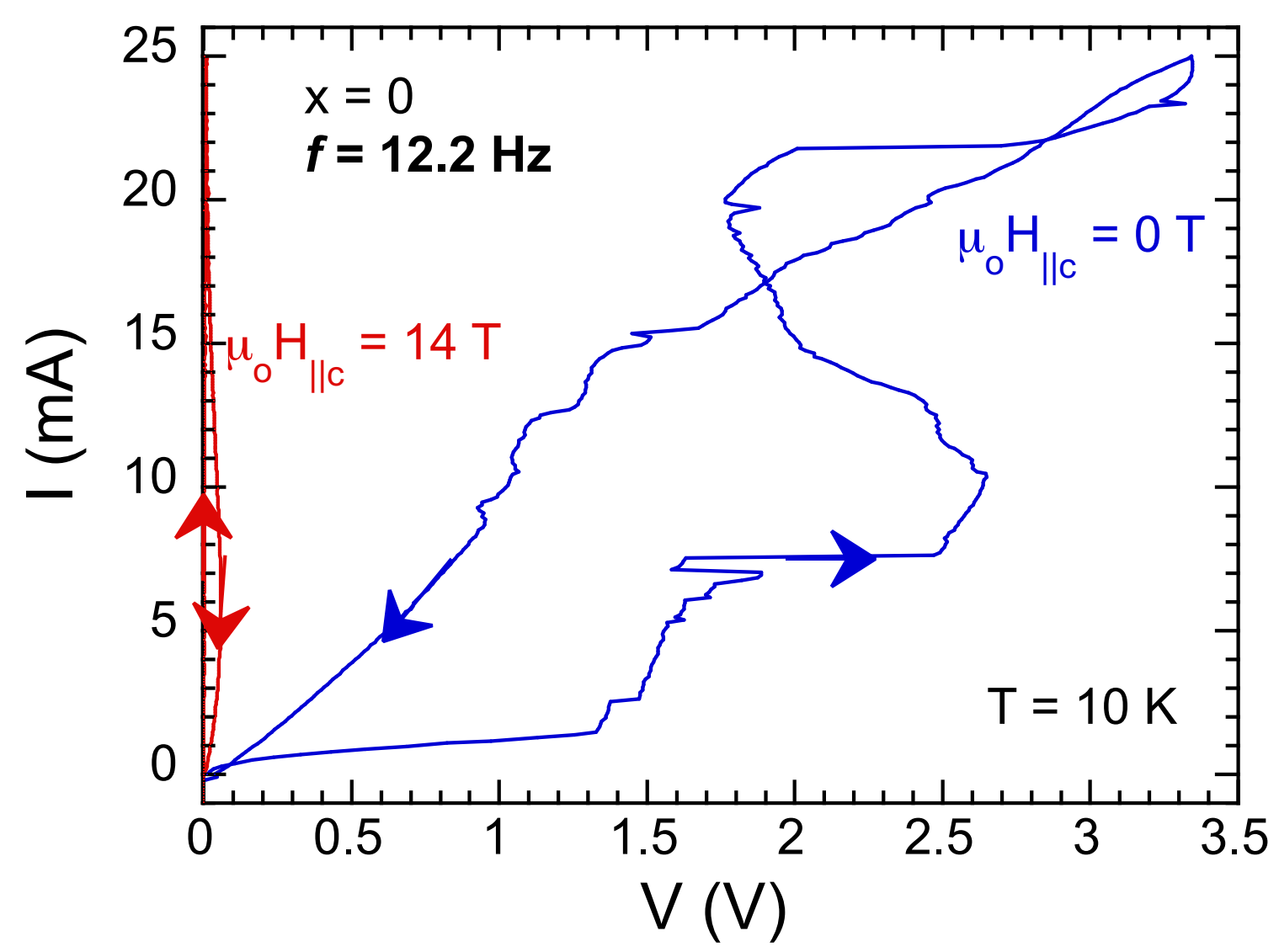

SFigure 3

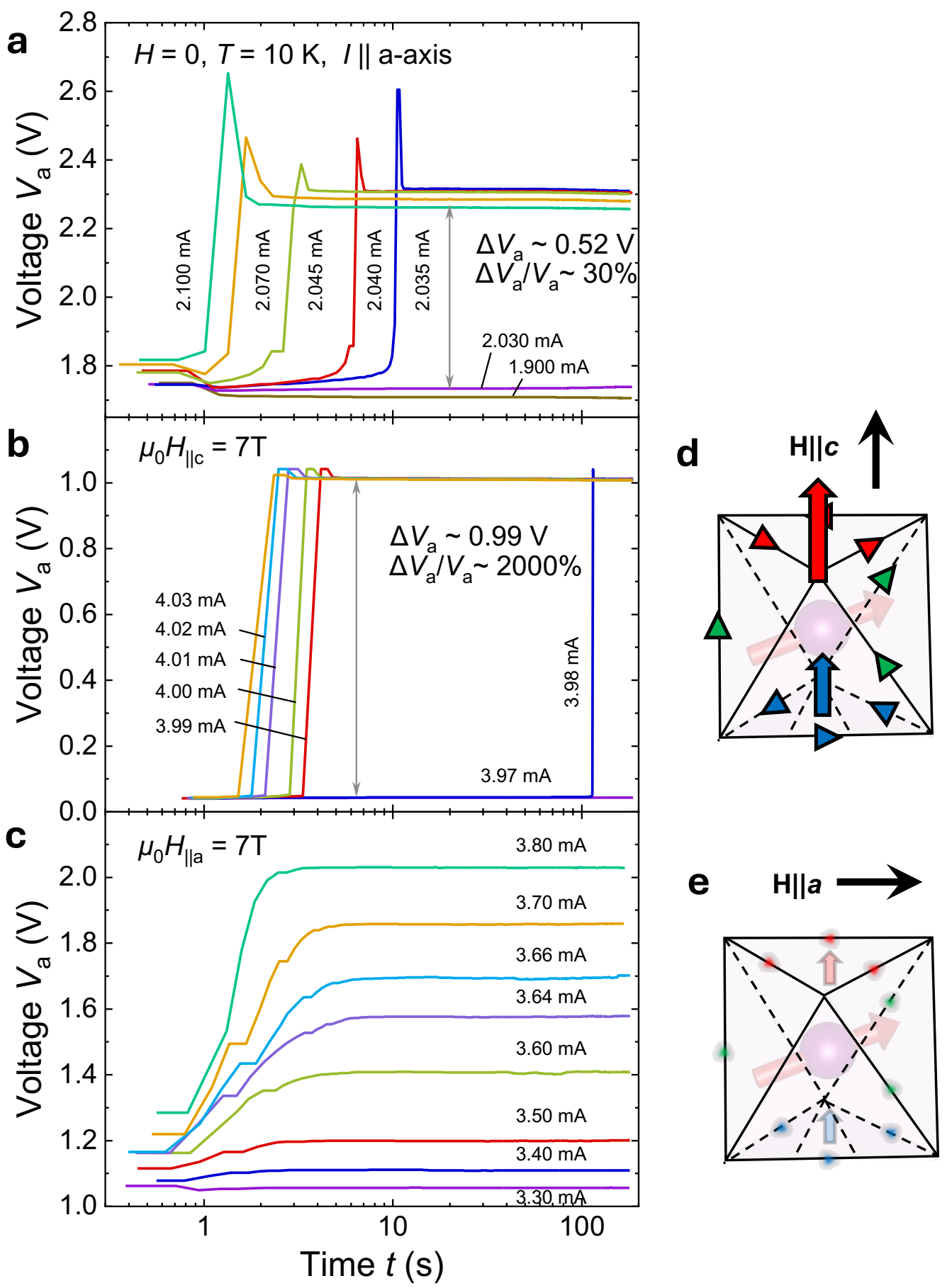

a
$H = 0$, $T = 10$ K, $I$ || a-axis
2.100 mA
2.070 mA
2.045 mA
2.040 mA
2.035 mA
$\Delta V_a \sim 0.52$ V
$\Delta V_a/V_a \sim 30\%$
2.030 mA
1.900 mA
b
$\mu_0 H_{||c} = 7$T
$\Delta V_a \sim 0.99$ V
$\Delta V_a/V_a \sim 2000\%$
4.03 mA
4.02 mA
4.01 mA
4.00 mA
3.99 mA
3.98 mA
3.97 mA
c
$\mu_0 H_{||a} = 7$T
3.80 mA
3.70 mA
3.66 mA
3.64 mA
3.60 mA
3.50 mA
3.40 mA
3.30 mA
Voltage $V_a$ (V)
Time $t$ (s)
d
H||*c*
e
H||*a*

SFigure 4

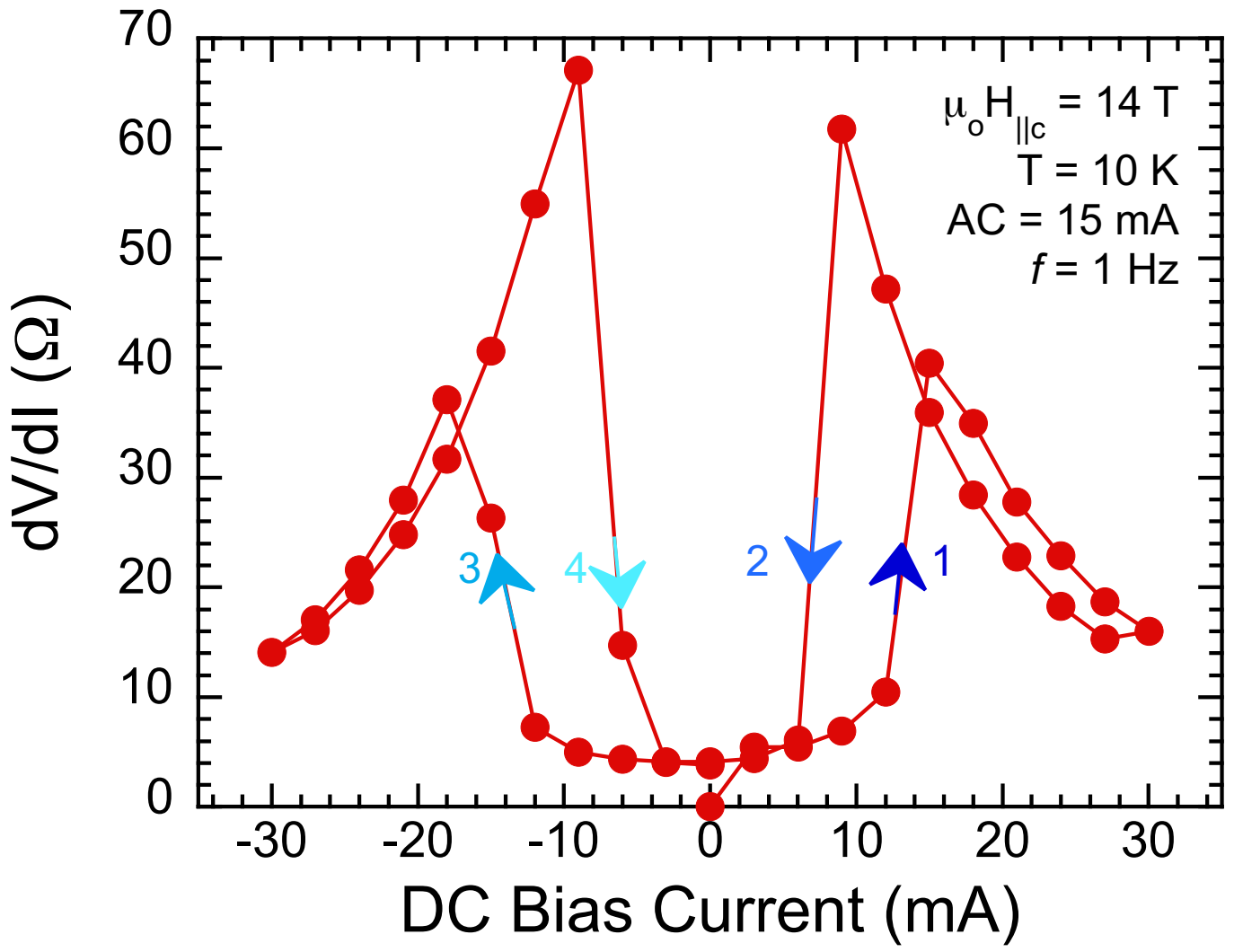

SFigure 5

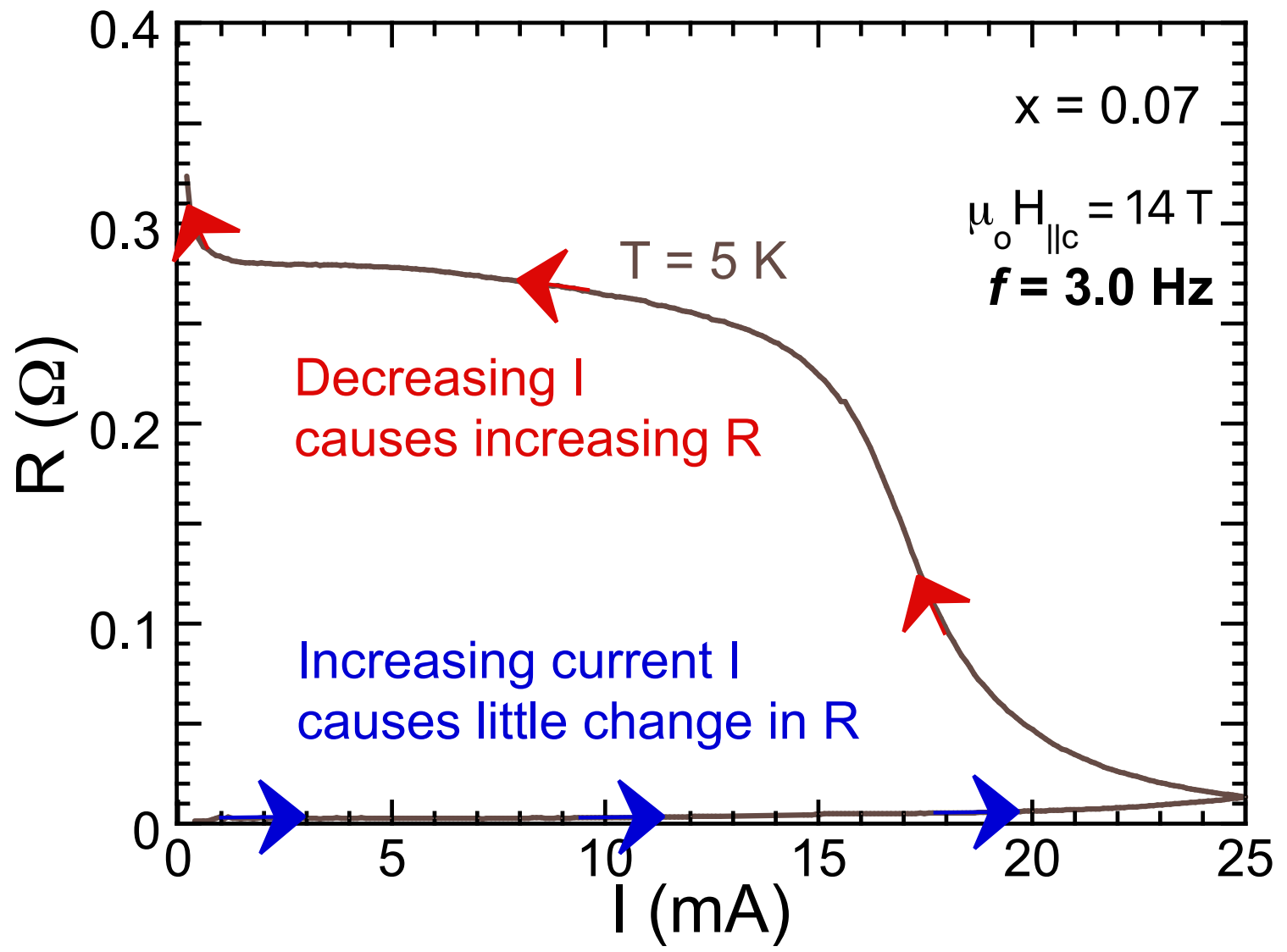

SFigure 6

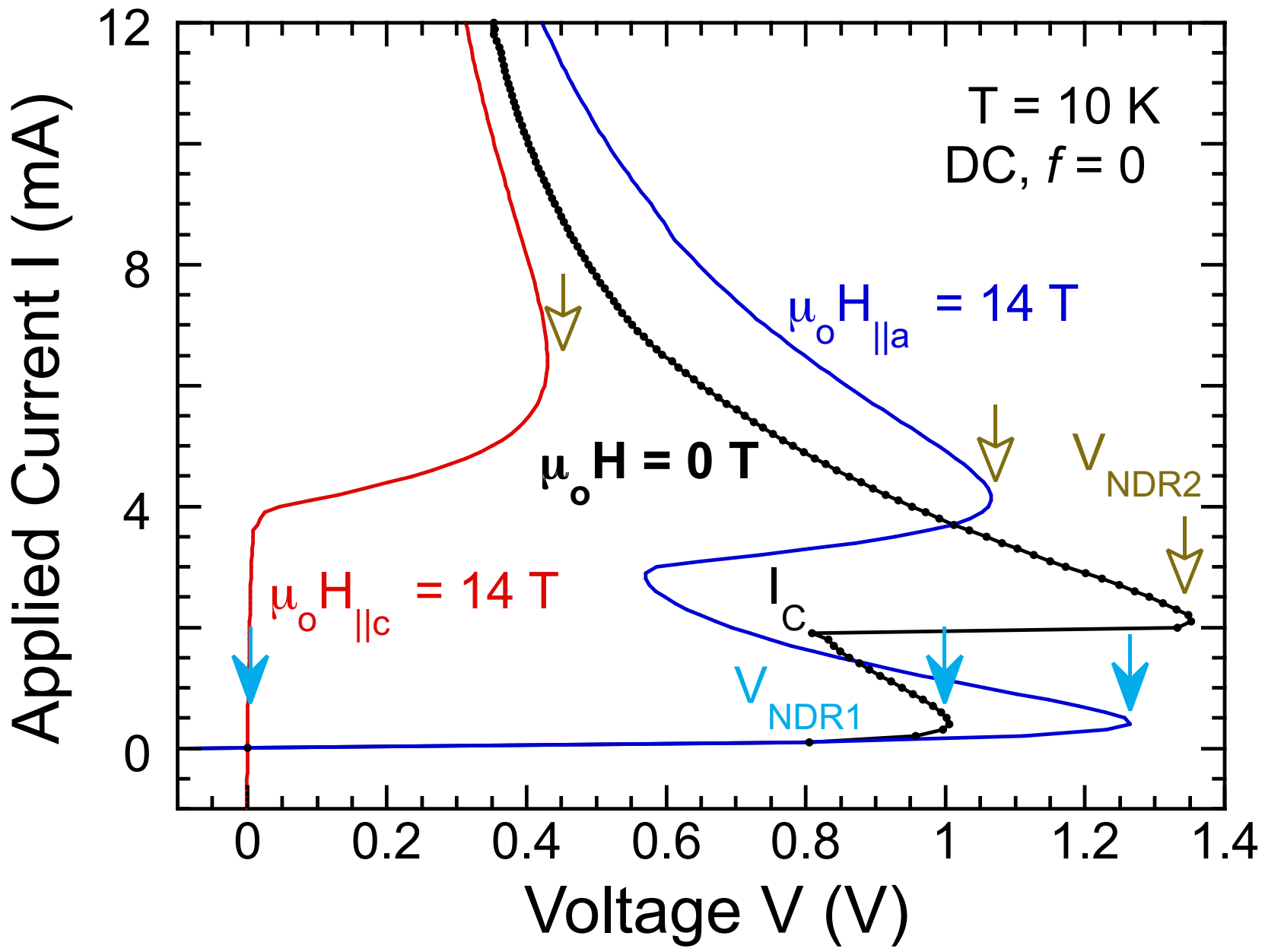

T = 10 K
DC, f = 0
$\mu_o H_{||a}$ = 14 T
$\mu_o H$ = 0 T
$\mu_o H_{||c}$ = 14 T
$V_{NDR2}$
$I_C$
$V_{NDR1}$
Applied Current I (mA)
Voltage V (V)

SFigure 7

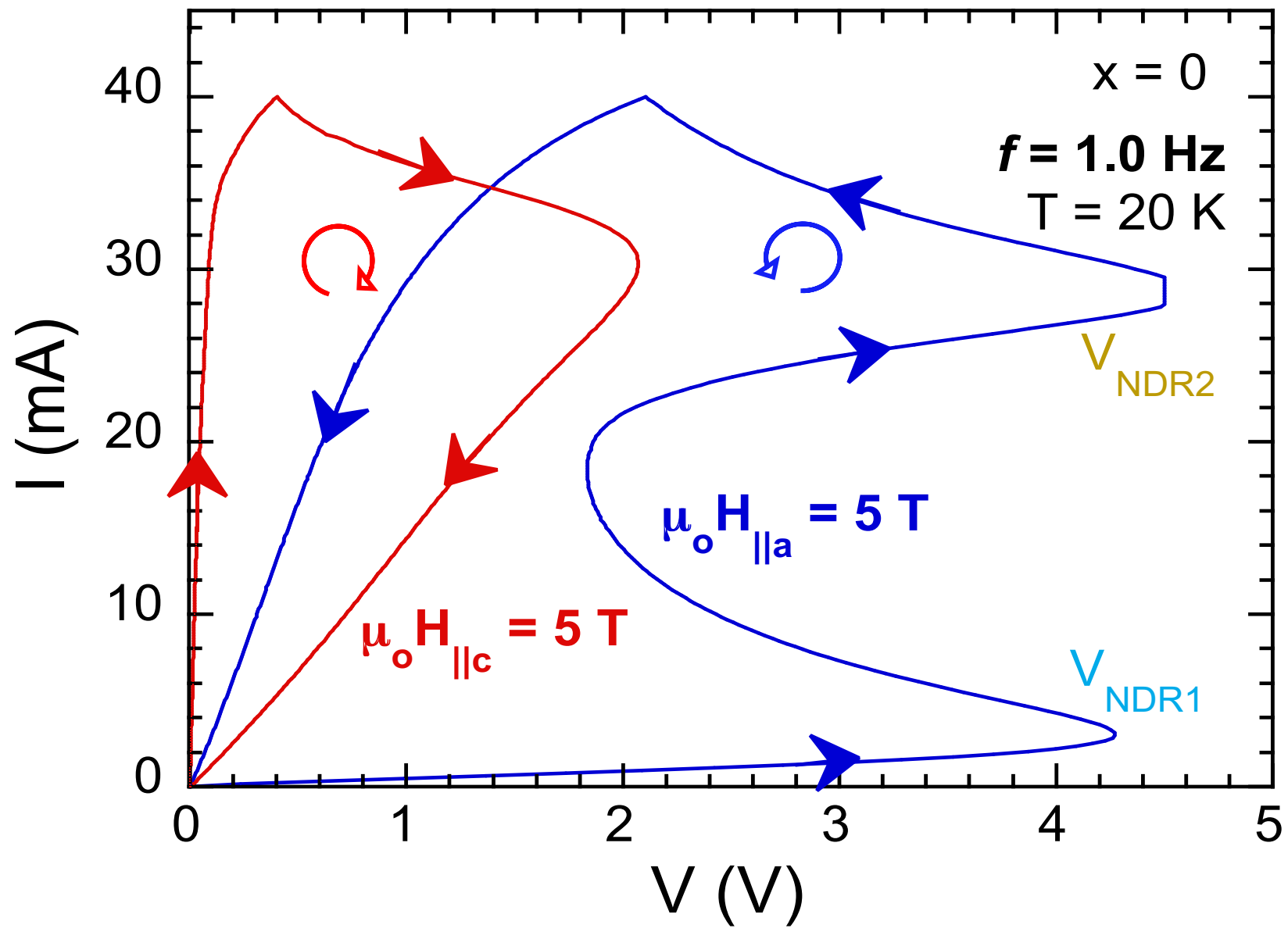

SFigure 8

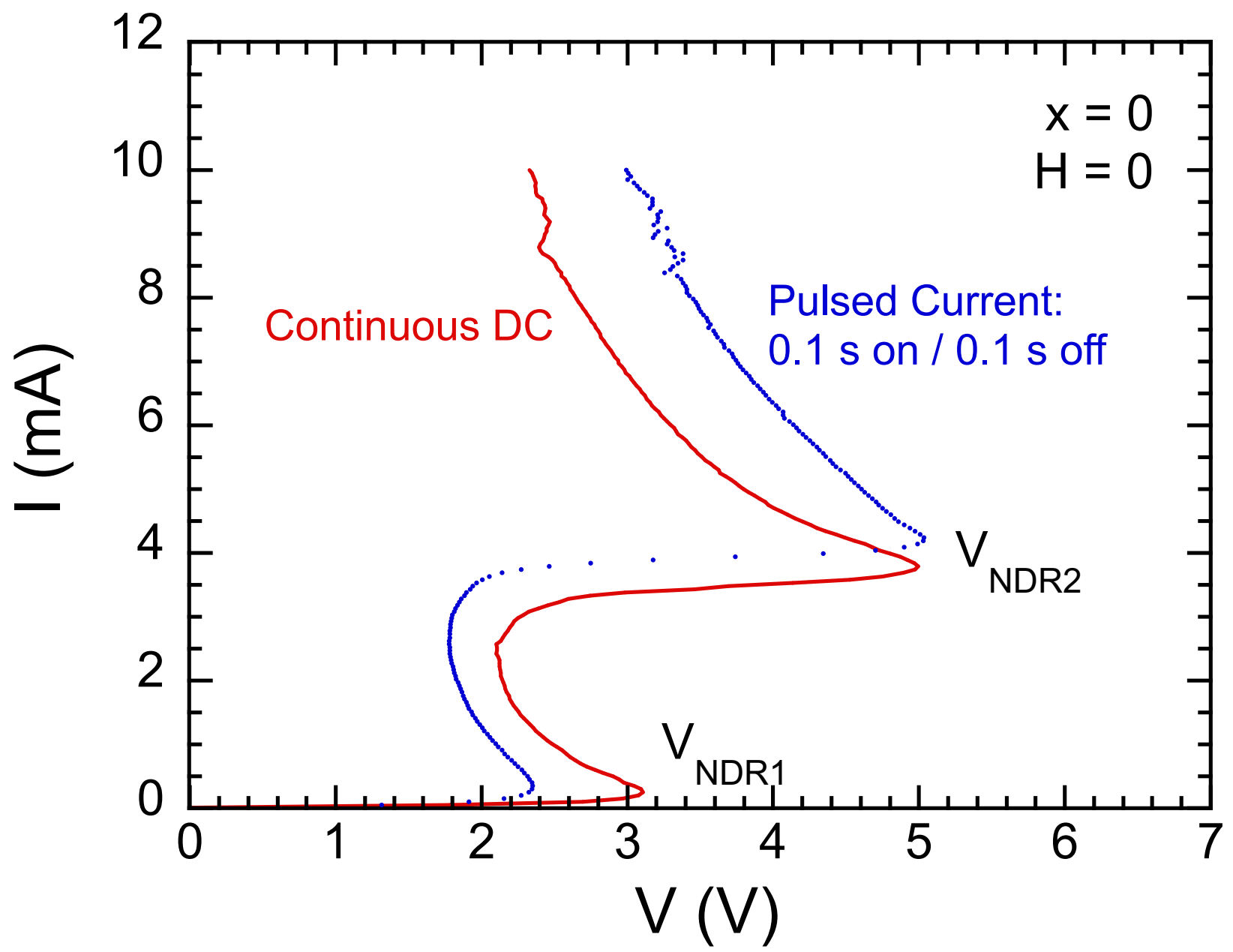

SFigure 9

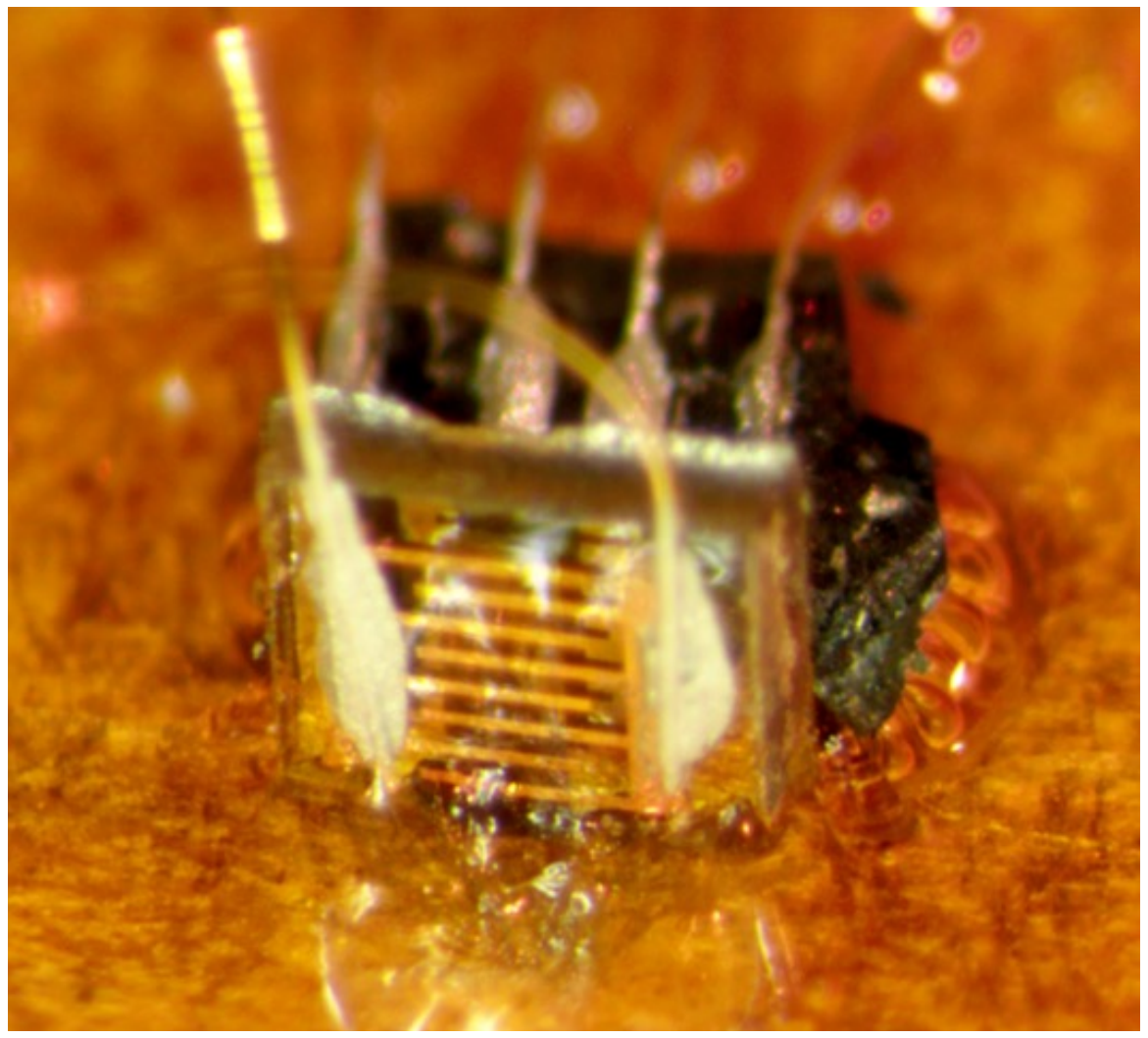

SFigure 10

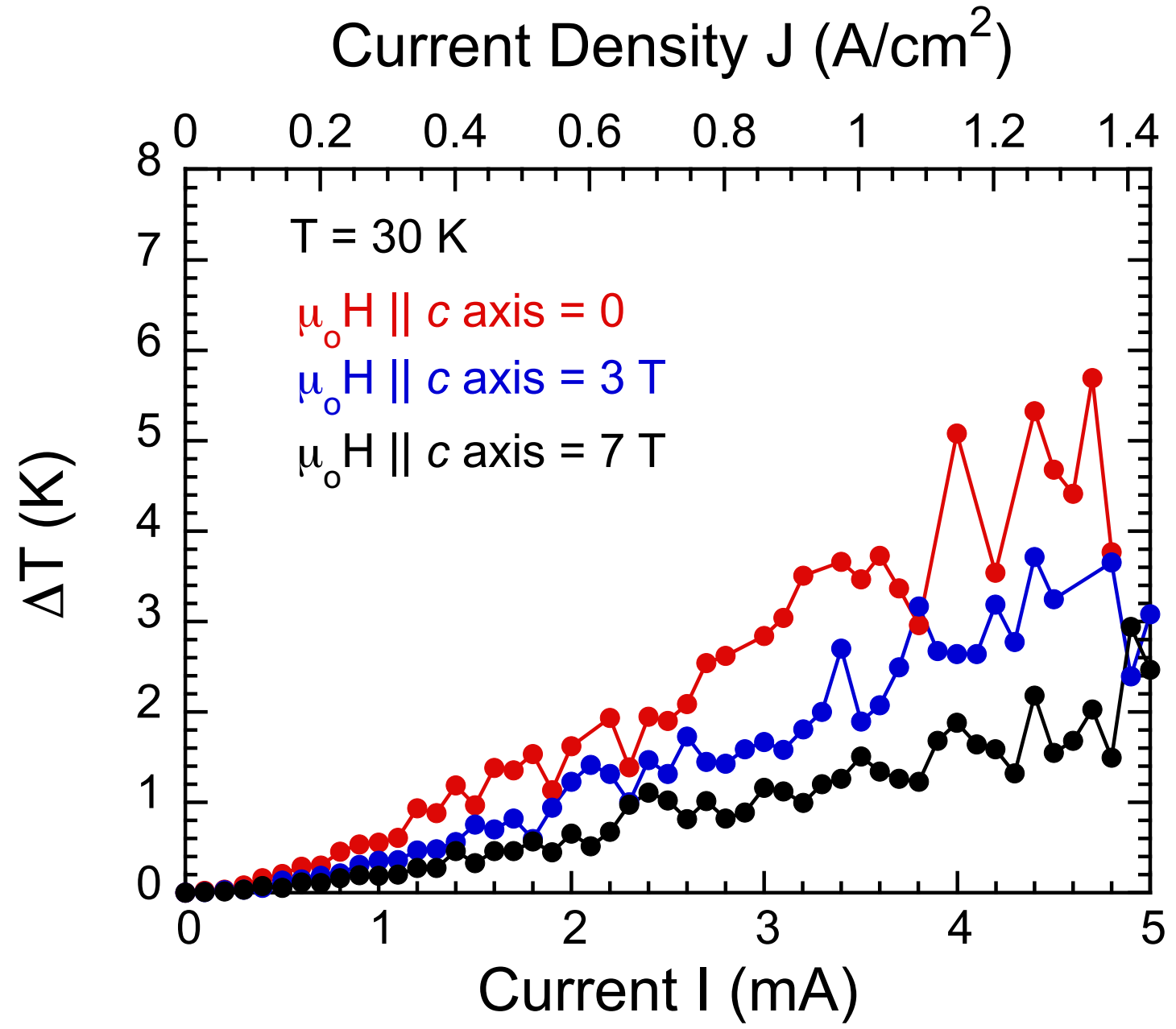

SFigure 11

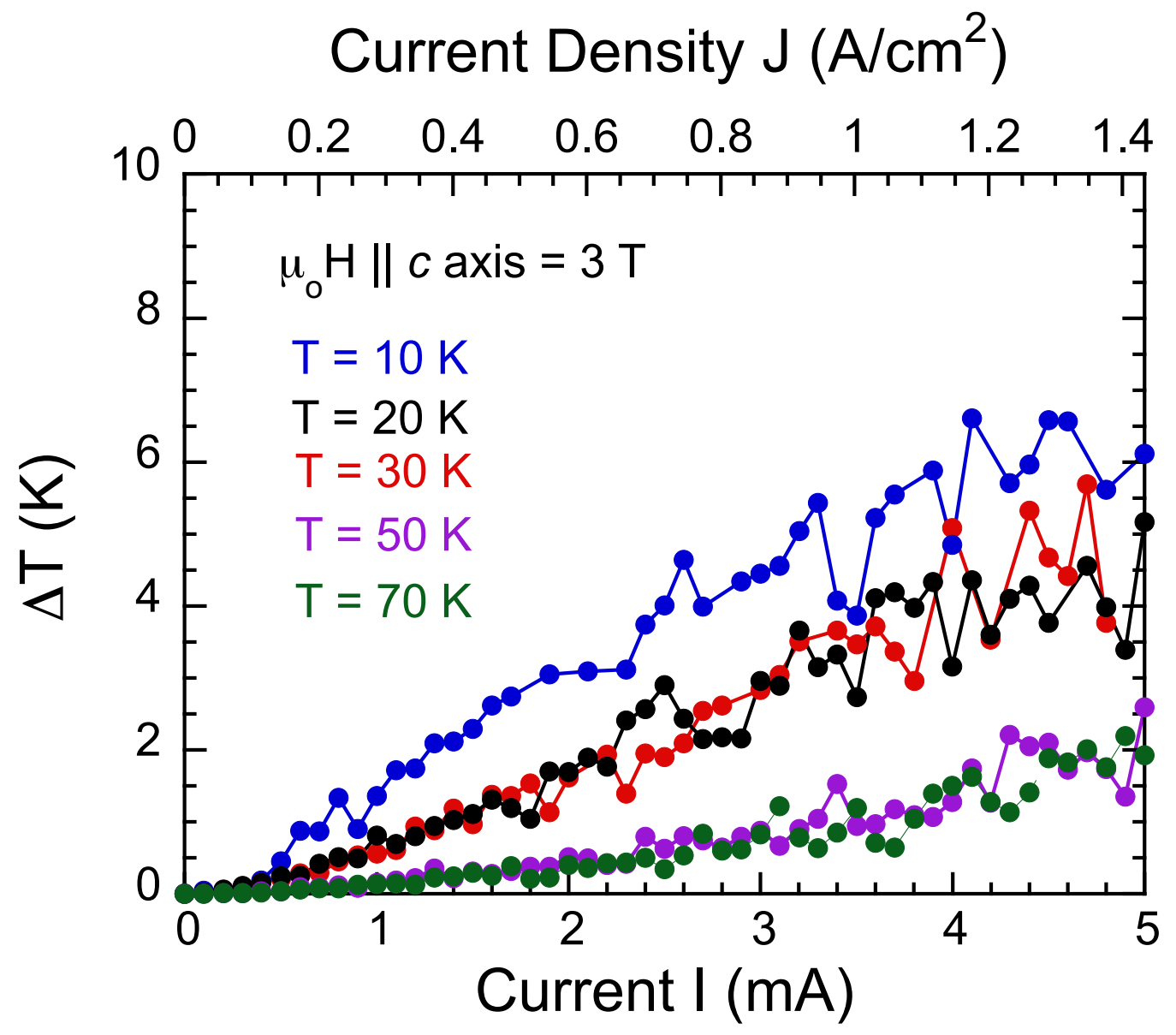
Current Density J (A/cm2)
0
0.2
0.4
0.6
0.8
1
1.2
1.4
10
8
6
4
2
0
ΔT (K)
μoH || c axis = 3 T
T = 10 K
T = 20 K
T = 30 K
T = 50 K
T = 70 K
0
1
2
3
4
5
Current I (mA)

SFigure 12